\documentstyle[numreferences,editedvolume,epsf,psfig,epsfig]{crckapb}

\begin{opening}
\title{RANDOM MATRIX THEORY AND DIRAC SPECTRUM AT NONZERO TEMPERATURE
       AND DENSITY \protect }

\author{B.A. BERG}
\institute{Department of Physics, The Florida State University,\\
                Tallahassee, FL 32306, USA}
\author{H. MARKUM}
\author{R. PULLIRSCH}
\institute{Institut f\"ur Kernphysik, Technische Universit\"at Wien, \\
                A-1040 Vienna, Austria}
\author{T. Wettig}
\institute{Department of Physics, Yale University,\\
                New Haven, CT 06520-8120, USA}

\end{opening}

\runningtitle{RANDOM MATRIX THEORY AND DIRAC SPECTRUM \ldots}

\makeatletter
\newbox\slashbox \setbox\slashbox=\hbox{\large$/$}
\def\pslash#1{\setbox\@tempboxa=\hbox{$#1$}
  \@tempdima=0.5\wd\slashbox \advance\@tempdima 0.5\wd\@tempboxa
  \copy\slashbox \kern-\@tempdima \box\@tempboxa}
\def\FMSlash{\protect\pslash}
\makeatother

\begin{document}


{ We investigate the eigenvalue spectrum of the staggered Dirac matrix
  in SU(3) gauge theory and in full QCD as well as in quenched U(1)
  theory on various lattice sizes.  As a measure of the fluctuation
  properties of the eigenvalues, we study the nearest-neighbor spacing
  distribution, $P(s)$.  
  We further study lattice QCD at nonzero chemical potential, $\mu$, by
  constructing the spacing distribution of adjacent eigenvalues in the
  complex plane.  We find that in all regions of their phase diagrams,
  compact lattice gauge theories have bulk spectral correlations given
  by random matrix theory, which is an indication for quantum chaos.
%
} 

\section{QCD at Nonzero Temperature}

The properties of the eigenvalues of the Dirac operator are of great
importance for the understanding of certain features of QCD.  For
example, the accumulation of small eigenvalues is, via the
Banks-Casher formula \cite{Bank80}, related to the spontaneous
breaking of chiral symmetry.  Recently, the fluctuation properties of
the eigenvalues in the bulk of the spectrum have also attracted
attention.  It was shown in Ref.~\cite{Hala95} that on the scale of
the mean level spacing they are described by random matrix theory
(RMT). In particular, the nearest-neighbor spacing distribution
$P(s)$, i.e., the distribution of spacings $s$ between adjacent
eigenvalues on the unfolded scale, agrees with the Wigner
surmise of RMT.  According to the Bohigas-Giannoni-Schmit conjecture
\cite{Bohi84}, quantum systems whose classical counterparts are
chaotic have a nearest-neighbor spacing distribution given by RMT
whereas systems whose classical counterparts are integrable obey a
Poisson distribution, $P_{\rm P}(s)=e^{-s}$.  Therefore, the specific
form of $P(s)$ is often taken as a criterion for the presence or
absence of ``quantum chaos''.

In RMT, one has to distinguish several universality classes which are
determined by the symmetries of the system.  For the case of the QCD
Dirac operator, this classification was done in Ref.~\cite{Verb94}.
Depending on the number of colors and the representation of the
quarks, the Dirac operator is described by one of the three chiral
ensembles of RMT.  As far as the fluctuation properties in the bulk of
the spectrum are concerned, the predictions of the chiral ensembles
are identical to those of the ordinary ensembles \cite{Fox64}.  In
Ref.~\cite{Hala95}, the Dirac matrix was studied in SU(2) using both
staggered and Wilson fermions which correspond to the chiral
symplectic and orthogonal ensemble, respectively.  Here \cite{Pull98},
we study SU(3) with staggered fermions which corresponds to the chiral
unitary ensemble. The RMT result for the
nearest-neighbor spacing distribution can
be expressed in terms of so-called prolate spheroidal functions, see
Ref.~\cite{Meht91}.  A very good
approximation to $P(s)$ is provided by
the Wigner surmise for the unitary ensemble,
\begin{equation} \label{wigner}
  P_{\rm W}(s)=\frac{32}{\pi^2}s^2e^{-4s^2/\pi} \:.
\end{equation}

We generated gauge field configurations using the standard Wilson
plaquette action for SU(3) with and without dynamical fermions in the
Kogut-Susskind prescription. We have worked on a $6^3\times 4$ lattice
with various values of the inverse gauge coupling $\beta=6/g^2$ both
in the confinement and deconfinement phase.  We typically produced 10
independent equilibrium configurations for each $\beta$.  Because of
the spectral ergodicity property of RMT one can replace ensemble
averages by spectral averages if one is only interested in bulk
properties.

The Dirac operator, $\FMSlash{D}=\FMSlash{\partial}+ig\FMSlash{A}$, is
anti-hermitian so that all eigenvalues are imaginary.  For
convenience, we denote them by $i\lambda_n$ and refer to the
$\lambda_n$ as the eigenvalues.  Because of
$\{\FMSlash{D},\gamma_5\}=0$ the nonzero $\lambda_n$ occur in pairs of
opposite sign.  All spectra were checked against the analytical sum
rules $\sum_{n} \lambda_n = 0$ and $\sum_{\lambda_n>0} \lambda_n^2 =
3V$, where V is the lattice volume.  To construct the nearest-neighbor
spacing distribution from the eigenvalues, one first has to
``unfold'' the spectra \cite{Meht91}.

\begin{figure}
\begin{center}
\begin{tabular}{ccccc}
  & {\large Confinement $\beta=5.2$}  & \hspace*{10mm}   & &
  {\large Deconfinement $\beta=5.4$} \\
  \vspace*{0mm}
  & $ma=0.05$ &&& $ma=0.05$ \\[2mm]
  \multicolumn{2}{c}{\epsfxsize=5cm\epsffile{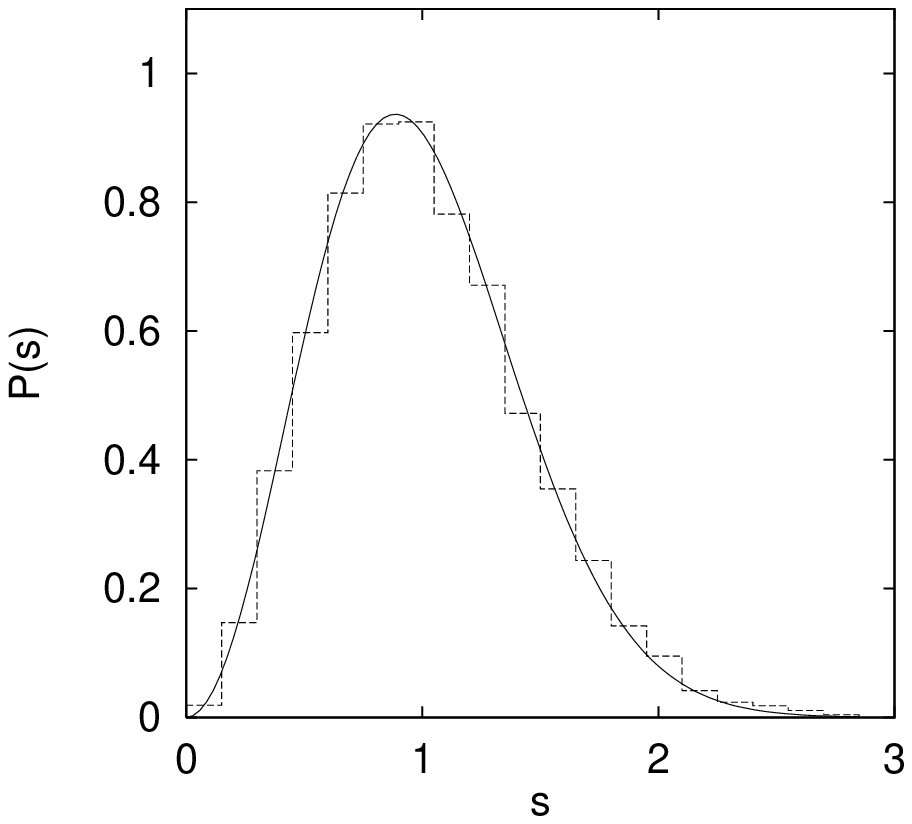}} &&
  \multicolumn{2}{c}{\epsfxsize=5cm\epsffile{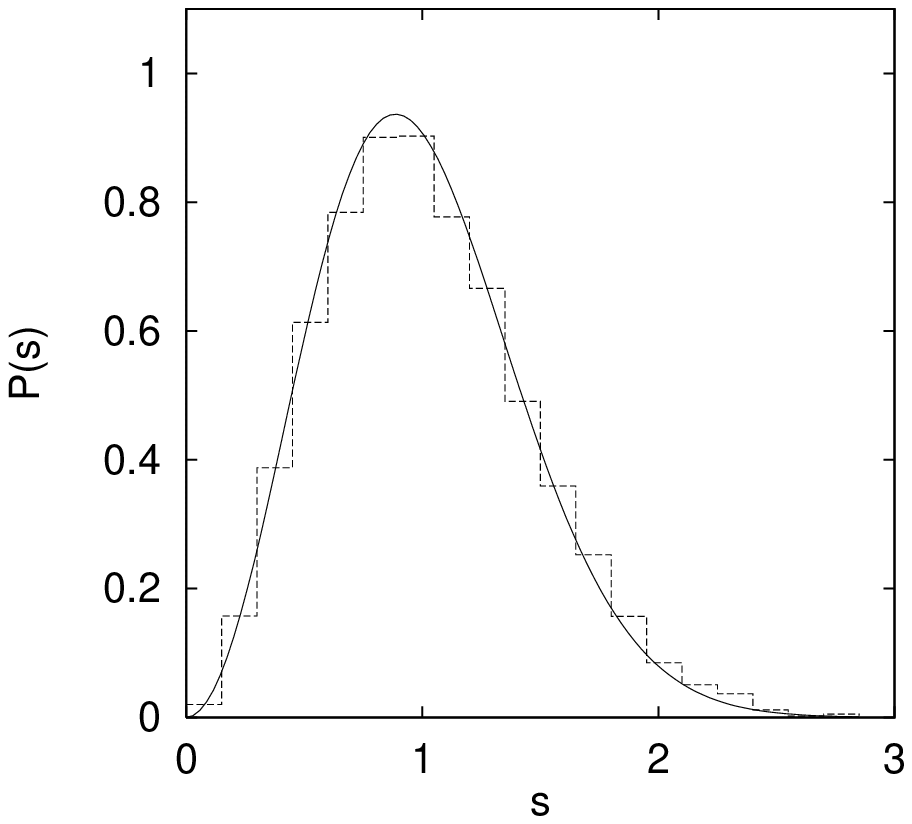}}
\end{tabular}
\end{center}
\vspace*{-3mm}
\caption{The nearest-neighbor spacing distribution $P(s)$ on a $6^3 
  \times 4$ lattice in full QCD (histograms) compared with the
  random matrix result (solid lines). There are no changes in $P(s)$
  across the deconfinement phase transition.}
\label{fintemp}
\end{figure}

Figure~\ref{fintemp} compares $P(s)$ of full QCD with $N_f = 3$
flavors and quark mass $ma=0.05$ to the RMT result.  In the confinement
as well as in the deconfinement phase we observe agreement with RMT up
to very high $\beta$ (not shown).  The observation that $P(s)$ is not
influenced by the presence of dynamical quarks could have been
expected from the results of Ref.~\cite{Fox64}, which
apply to the case of massless quarks. Our
results, and those of~\cite{Hala95}, indicate that massive
dynamical quarks do not affect $P(s)$ either.

\begin{figure*}[t]
  \centerline{\psfig{figure=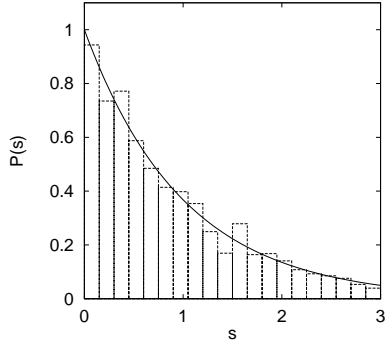,width=5cm}}
  \caption{Nearest-neighbor spacing distribution $P(s)$ for the free
            Dirac operator on a $53\times 47\times 43\times 41$ lattice
            compared with a Poisson distribution, $e^{-s}$.}
  \label{free}
\end{figure*}

No signs for a transition to Poisson regularity are found. The
deconfinement phase transition does not seem to coincide with a
transition in the spacing distribution. For very large values of
$\beta$ far into the deconfinement region, the eigenvalues
start to approach the degenerate eigenvalues of the free theory, given
by $\lambda^2=\sum_{\mu=1}^4 \sin^2(2\pi n_\mu/L_\mu)/a^2$, where $a$
is the lattice constant, $L_{\mu}$ is the number of lattice sites in
the $\mu$-direction, and $n_\mu=0,\ldots,L_\mu-1$.  In this case, the
spacing distribution is neither Wigner nor Poisson.
It is possible to lift the degeneracies of the free
eigenvalues using an asymmetric lattice where $L_x$, $L_y$, etc. are
relative primes and, for large lattices, the distribution 
is then Poisson, $P_{\rm P}(s)=e^{-s}$, see Fig.~\ref{free}.

\section{QCD at Nonzero Density}

Physical systems which are described by non-hermitian operators have
attracted a lot of attention recently, among others QCD at nonzero
chemical potential $\mu$~\cite{Step96}. There, the Dirac operator
loses its hermiticity properties so that its eigenvalues become
complex. The aim of the present analysis is to investigate whether
non-hermitian RMT is able to describe the fluctuation properties of
the complex eigenvalues of the QCD Dirac operator.  The eigenvalues
are generated on the lattice for various values of $\mu$.  We apply a
two-dimensional unfolding procedure to separate the average eigenvalue
density from the fluctuations and construct the nearest-neighbor
spacing distribution, $P(s)$, of adjacent eigenvalues in the complex
plane.  The data are then compared to analytical predictions of
non-hermitian RMT.

We start with a few definitions. A formulation of the QCD Dirac
operator at $\mu\ne0$ on the lattice in the staggered scheme is
given by \cite{Hase83}
\begin{eqnarray}
  \label{Dirac}
  M_{x,y}(U,\mu) & = &
  \frac{1}{2a} \sum\limits_{\nu=\hat{x},\hat{y},\hat{z}}
  \left[U_{\nu}(x)\eta_{\nu}(x)\delta_{y,x\!+\!\nu}-{\rm h.c.}\right]
  \nonumber\\
  &&
  + \, \frac{1}{2a}\left[U_{\hat{t}}(x)\eta_{\hat{t}}(x)e^{\mu}
    \delta_{y,x\!+\!\hat{t}}
    -U_{\hat{t}}^{\dagger}(y)\eta_{\hat{t}}(y)
    e^{-\mu}\delta_{y,x\!-\!\hat{t}}\right]
\end{eqnarray}
with the link variables $U$ and the staggered phases $\eta$.

We consider the gauge group SU(3) which corresponds to the symmetry 
class of the chiral unitary ensemble of RMT \cite{Verb94,Hala97a}.
At zero chemical
potential, all Dirac eigenvalues are purely imaginary, and the
nearest-neighbor spacing distribution, $P(s)$, of the lattice data
agrees with the Wigner surmise of hermitian RMT,
Eq.~(\protect\ref{wigner}), both in the confinement and in the
deconfinement phase (see~Fig.~\ref{fintemp}).
This finding implies strong correlations of the eigenvalues. 
For uncorrelated eigenvalues $P(s)$ is given by the Poisson
distribution.

For a complex spectrum, we define $P(s)$ to represent the spacing
distribution of nearest neighbors in the complex plane, i.e., for each
eigenvalue $z_0$ one identifies the eigenvalue $z_1$ for which
$s=|z_1-z_0|$ is a minimum \cite{Grob88}.  
After ensemble averaging, one obtains a function $P(s,z_0)$
which, in general, depends on $z_0$.  The dependence on $z_0$
can be eliminated by unfolding the spectrum, i.e., by applying a local
rescaling of the energy scale so that the average spectral density is
constant in a bounded region in the complex plane and zero outside
\cite{Mark99}.  After unfolding, a spectral average
over $z_0$ yields $P(s)$.  

For $\mu>0$, the eigenvalues of the matrix in Eq.~(\ref{Dirac}) move
into the complex plane.  If the real and imaginary parts of the
strongly correlated eigenvalues have approximately the same average
magnitude, the system should be described by the Ginibre ensemble of
non-hermitian RMT \cite{Gini65}.  
In the Ginibre ensemble, the average spectral density is already
constant inside a circle and zero outside, respectively.
In this case, unfolding is not necessary, and $P(s)$ is given by
\cite{Grob88}
\begin{equation} \label{Ginibre}
  P_{\rm G}(s)  =  c\, p(cs)\:, ~~p(s) = 
  2s\lim_{N\to\infty}\left[\prod_{n=1}^{N-1}e_n(s^2)\,e^{-s^2}
  \right] \sum_{n=1}^{N-1}\frac{s^{2n}}{n!e_n(s^2)}\:,
\end{equation}
where $e_n(x)=\sum_{m=0}^n x^m/m!$ and $c=\int_0^\infty ds \, s \,
p(s)=1.1429...$.  This result holds for strongly non-hermitian
matrices, i.e., for ${\rm Re}(z)\approx{\rm Im}(z)$ on average. In the
regime of weak non-hermiticity \cite{Fyod96}, where the typical
magnitude of the imaginary parts of the eigenvalues is equal to the
mean spacing of the real parts, the RMT prediction deviates from
Eq.~(\ref{Ginibre}).  
We shall
comment on this regime below.  For uncorrelated eigenvalues in the
complex plane, the Poisson distribution becomes \cite{Grob88}
\begin{equation}
  \label{Poisson}
  P_{\bar{\rm P}}(s)=\frac{\pi}{2}\,s\,e^{-\pi s^2/4}\:.
\end{equation}
This should not be confused with the Wigner distribution~(\ref{wigner}).

Our 
simulations were done with gauge group SU(3) on a $6^3\times4$ lattice
using $\beta=6/g^2=5.2$ in the confinement region and $\beta=5.4$ in
the deconfinement region for $N_f=3$ flavors of staggered fermions of
mass $ma=0.1$.    Despite major efforts \cite{Barb97} there
is currently no feasible solution to the problem of a complex weight
function in lattice simulations.   (In a random matrix model, the
numerical effort to generate a statistically significant ensemble of
configurations including the complex Dirac determinant was shown to
grow exponentially with $\mu^2N$, where $N$ is the lattice size
\cite{Hala97}.)  Therefore, the gauge field configurations were generated at
$\mu=0$, and the chemical potential was added to the Dirac matrix
afterwards. Both in the confinement and deconfinement, we sampled 50
independent configurations. 

\begin{figure}[-t]
  \begin{center}
    \epsfig{figure=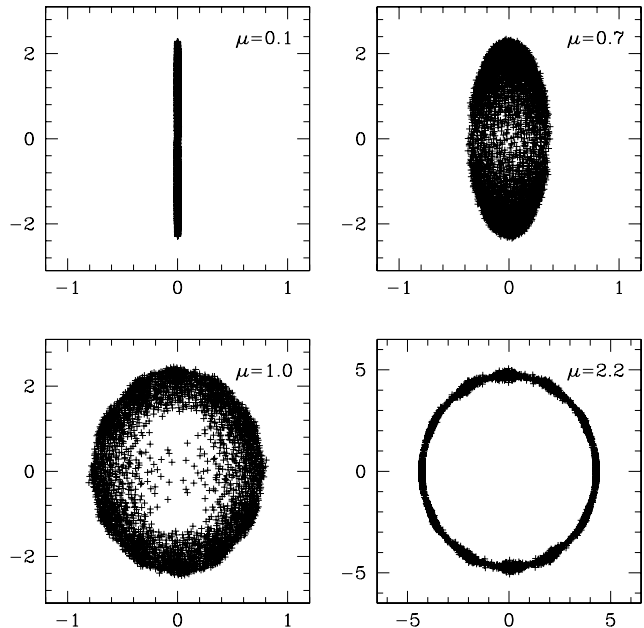,width=20pc}
  \end{center}
  \caption{Scatter plot of the eigenvalues of the Dirac operator (in
    units of $1/a$) in the complex plane at various values of $\mu$
    for a typical configuration of full QCD (generated at $\mu=0$) in
    the confinement region at $\beta=5.2$.}
  \label{spectra}
\end{figure}

Typical eigenvalue spectra are shown in Fig.~\ref{spectra} for four
different values of $\mu$ (in units of $1/a$) at $\beta=5.2$.  As
expected, the size of the real parts of the eigenvalues grows with
$\mu$, consistent with Ref.~\cite{Barb86}.  
Since the average spectral density is not constant, we have to
apply the unfolding method defined in \cite{Mark99}. 

\begin{figure}
  \begin{center}
    \epsfig{figure=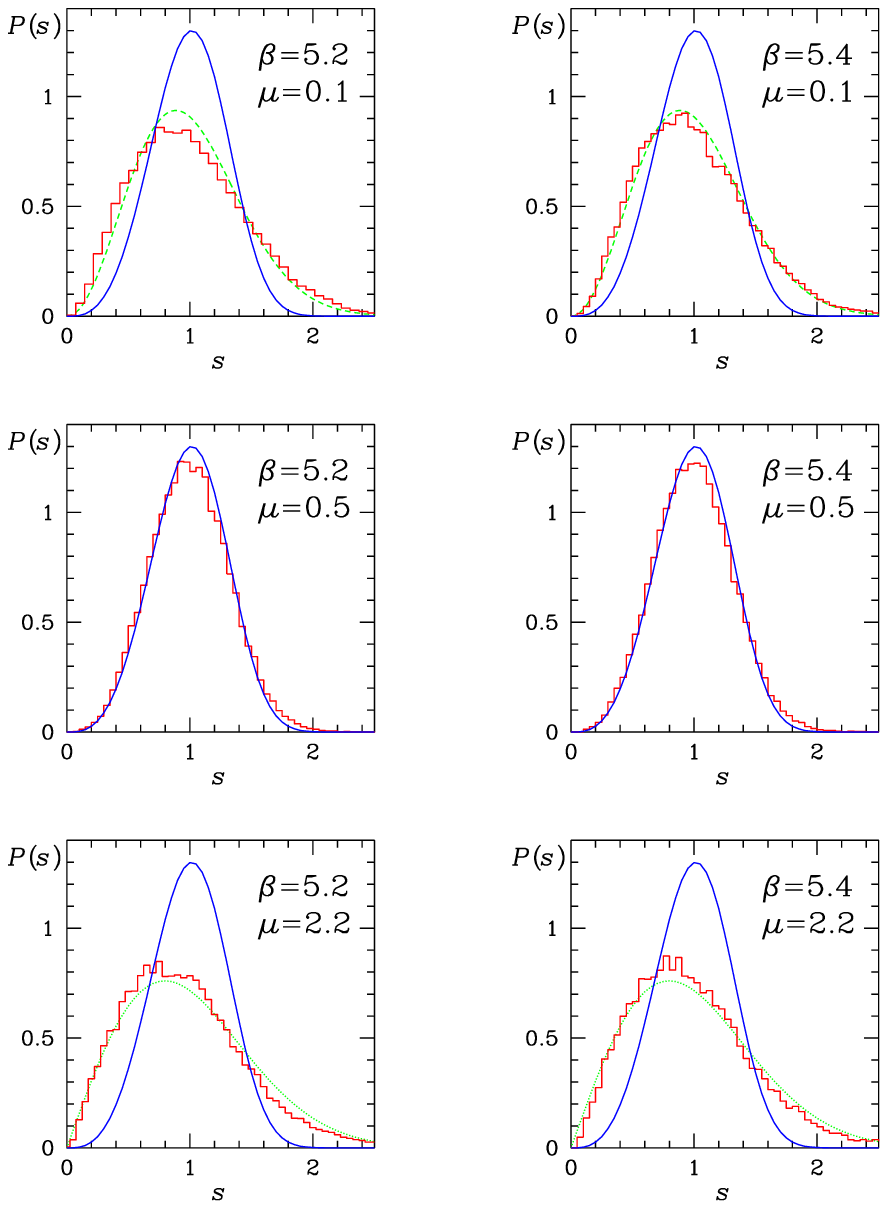}
  \end{center}
  \caption{Nearest-neighbor spacing distribution of the Dirac operator
    eigenvalues in the complex plane for various values of $\mu$ in
    the confinement (left) and deconfinement (right) phase. The
    histograms represent the lattice QCD data. The solid curve is
    the Ginibre distribution of Eq.~(\protect\ref{Ginibre}), the
    short-dashed curve in the first row the Wigner distribution of
    Eq.~(\protect\ref{wigner}), and the dotted curve in the last row
    the Poisson distribution of Eq.~(\protect\ref{Poisson}).}
  \vspace*{-3mm}
  \label{Psconf}
\end{figure}

Our results for $P(s)$ are presented in Fig.~\ref{Psconf}.  There are
minor quantitative but no qualitative differences between confinement
and deconfinement phase, which is consistent with our findings at
$\mu=0$ (see Fig.~\ref{fintemp}).  As a function of $\mu$, we expect
to find a transition from Wigner to Ginibre behavior in $P(s)$, as
is indeed seen in the figures.  For $\mu=0.1$, the data are still very
close to the Wigner distribution~(\ref{wigner}) whereas for
$0.5 \leq \mu \leq 0.7$ ($\mu=0.7$ not shown) we observe nice
agreement with the Ginibre distribution~(\ref{Ginibre}).
Values of $\mu$ in the crossover region between Wigner and Ginibre
behavior ($0.1<\mu<0.3$) correspond to the regime of weak
non-hermiticity mentioned above (the plots for $\mu=0.3$ can be found
in Ref.~\cite{Mark99}).  In this regime, the derivation of the spacing
distribution is a very difficult problem, and the only known
analytical result is $P(s,z_0)$ for small $s$, where $z_0$ is the
location in the complex plane (i.e., no unfolding is performed)
\cite{Fyod96}.  The small-$s$ behavior of Eqs.~(\ref{wigner}) and
(\ref{Ginibre}) is given by $P_{\rm W}(s)\propto s^2$ and $P_{\rm G}
(s)\propto s^3$, respectively, and in the regime of weak
non-hermiticity we have $P(s,z_0)\propto s^\alpha$ (for $s\ll1$) with
$2<\alpha<3$ \cite{Fyod96}.  This smooth crossover from $\alpha=2$ to
$\alpha=3$ is also observed in our unfolded data.

For $\mu > 0.7$ the lattice results for $P(s)$ deviate substantially
from the Ginibre distribution.  
The global spectral density of the lattice data for $\mu=1.0$ and 2.2
in Fig.~\ref{spectra} is very different from that of the Ginibre
ensemble.  This does not immediately imply that the local spectral
fluctuations are also different, but it is an indication for
qualitative changes.
The results for
$\mu=2.2$ in Fig.~\ref{Psconf} could be interpreted as Poisson
behavior, corresponding to uncorrelated eigenvalues. (In the hermitian
case at nonzero temperature, lattice simulations only show a
transition to Poisson behavior for $\beta\to\infty$ when the physical
box size shrinks and the theory becomes free \cite{Pull98}.) A plausible
explanation of the transition to Poisson behavior is provided by the
following two (related) observations. First, for large $\mu$ the terms
containing $e^\mu$ in Eq.~(\ref{Dirac}) dominate the Dirac matrix, giving
rise to uncorrelated eigenvalues. Second, for $\mu>1.0$ the fermion density
on the $6^3\times4$ lattice reaches saturation due to limited box size and
the Pauli exclusion principle.


\section{QED at Nonzero Temperature}

By now it is a well-known fact that the spectrum of the QCD Dirac
operator is related to universality classes of RMT, i.e., determined
by the global symmetries of the QCD partition function.
We have investigated 4d U(1) gauge theory which was not classified yet.
At $\beta_c \approx 1.01$ U(1) gauge theory undergoes a phase
transition between a confinement phase with mass gap and monopole
excitations for $\beta < \beta_c$ and the Coulomb phase which exhibits
a massless photon~\cite{BePa84} for $\beta > \beta_c$.
\begin{figure*}[t]
  \centerline{\psfig{figure=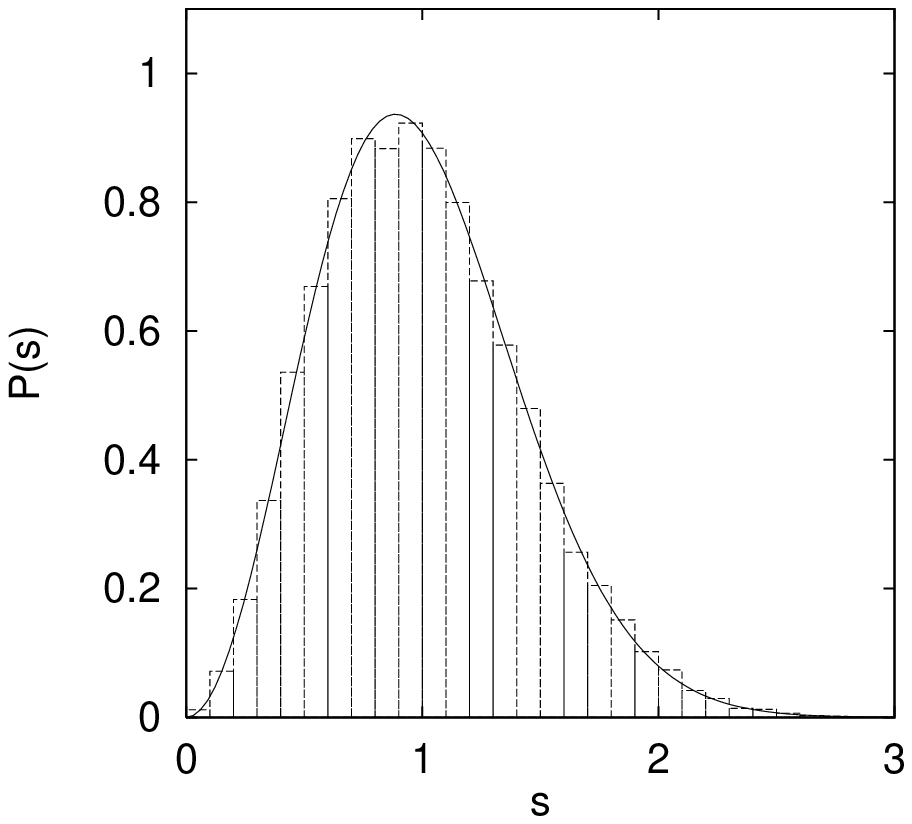,width=5cm}\hspace*{3mm}
    \psfig{figure=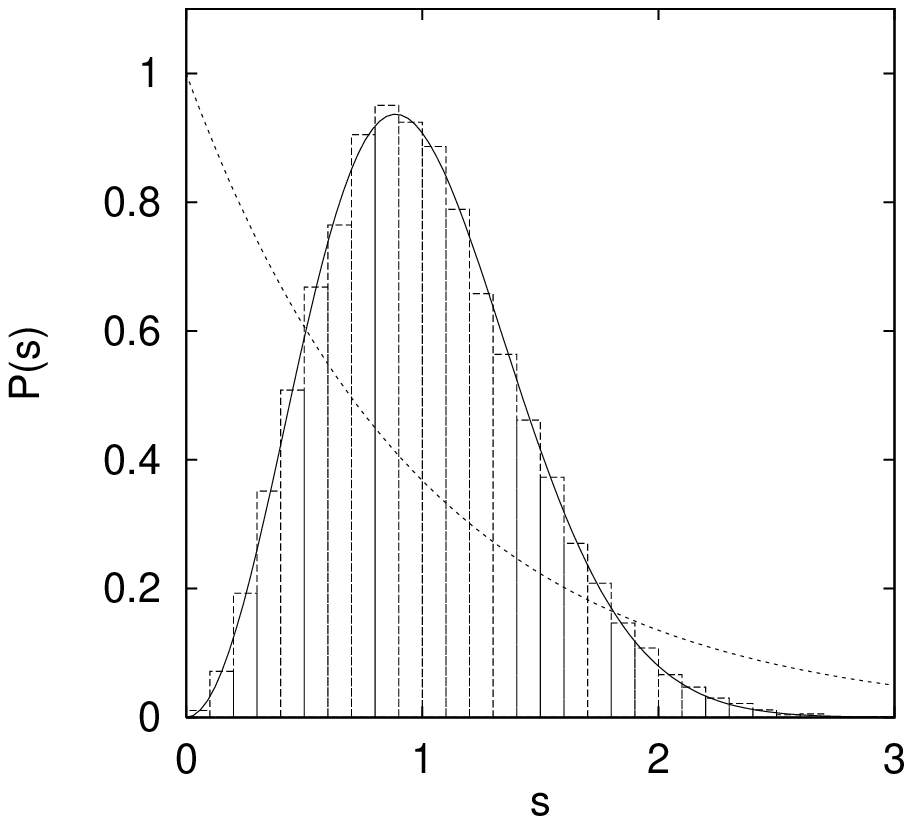,width=5cm}}
  \caption{Nearest-neighbor spacing distribution $P(s)$ for U(1) gauge
    theory on an $8^3\times 6$ lattice in the confined phase (left)
    and in the Coulomb phase (right). The theoretical curves are the chUE
    result, Eq.~(\ref{wigner}), and the Poisson distribution, $P_{\rm
      P}(s)=\exp(-s)$.}  \vspace*{-0.3cm}
  \label{f02}
\end{figure*}
As for SU(2) and SU(3) gauge groups, we expect the confined phase to
be described by RMT, whereas free fermions are known to yield the
Poisson distribution (see Fig.~\ref{free}). The question arose whether
the Coulomb phase will be described by RMT or by the Poisson
distribution \cite{BeMaPu99}.  The nearest-neighbor spacing
distributions for an $8^3\times 6$ lattice at $\beta=0.9$ (confined
phase) and at $\beta=1.1$ (Coulomb phase), averaged over 20
independent configuration, are depicted in Fig.~\ref{f02}. Both are
well described by the chiral unitary ensemble (chUE) of RMT.

We have continued the above investigation with a study of the
distribution of small eigenvalues in the confined phase. The
Banks-Casher formula~\cite{Bank80} relates the eigenvalue density
$\rho(\lambda)$ at $\lambda=0$ to the chiral condensate,
$ \Sigma = |\langle \bar{\psi} \psi \rangle| =
 \lim_{m\to 0}\lim_{V\to\infty} \pi\rho (0)/V$.
The microscopic spectral density,
$ \rho_s (z) = \lim_{V\to\infty}
 \rho \left( {z/V\Sigma } \right)/V\Sigma , $
should be given by the result for the chUE of RMT~\cite{ShVe92}.  This
function also generates the Leutwyler-Smilga sum rules~\cite{LeSm92}.

\begin{figure}[-t]
  \begin{center}
    \hspace*{10mm}\psfig{figure=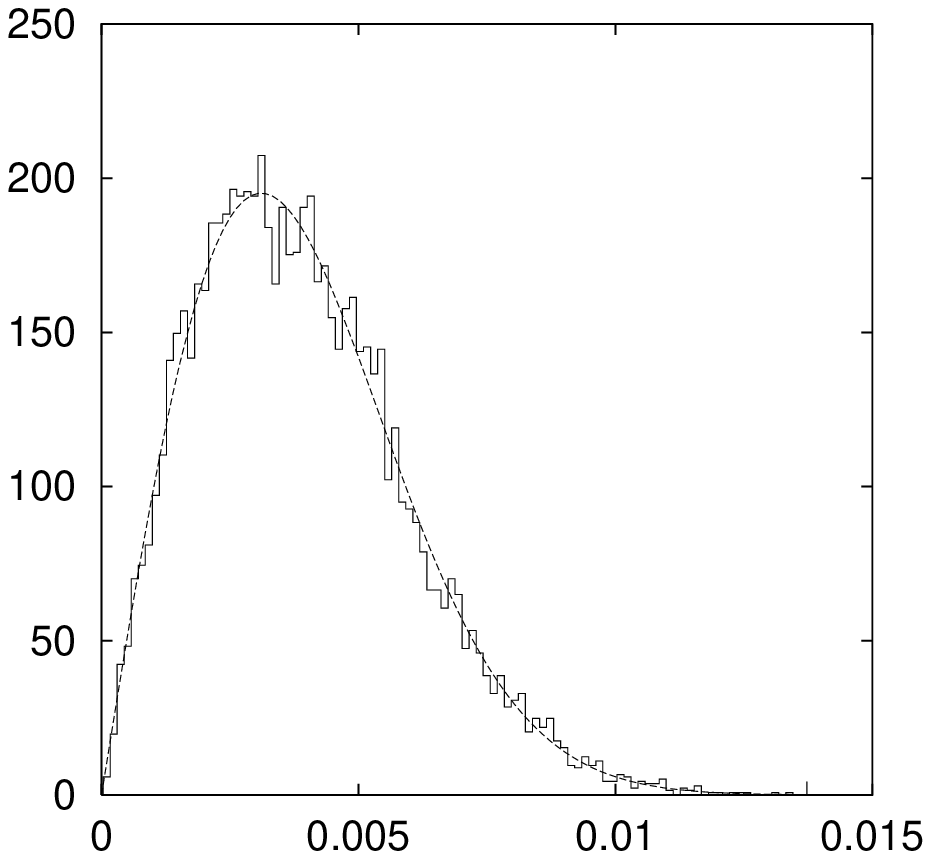,width=4.6cm}
    \hspace*{15mm}\psfig{figure=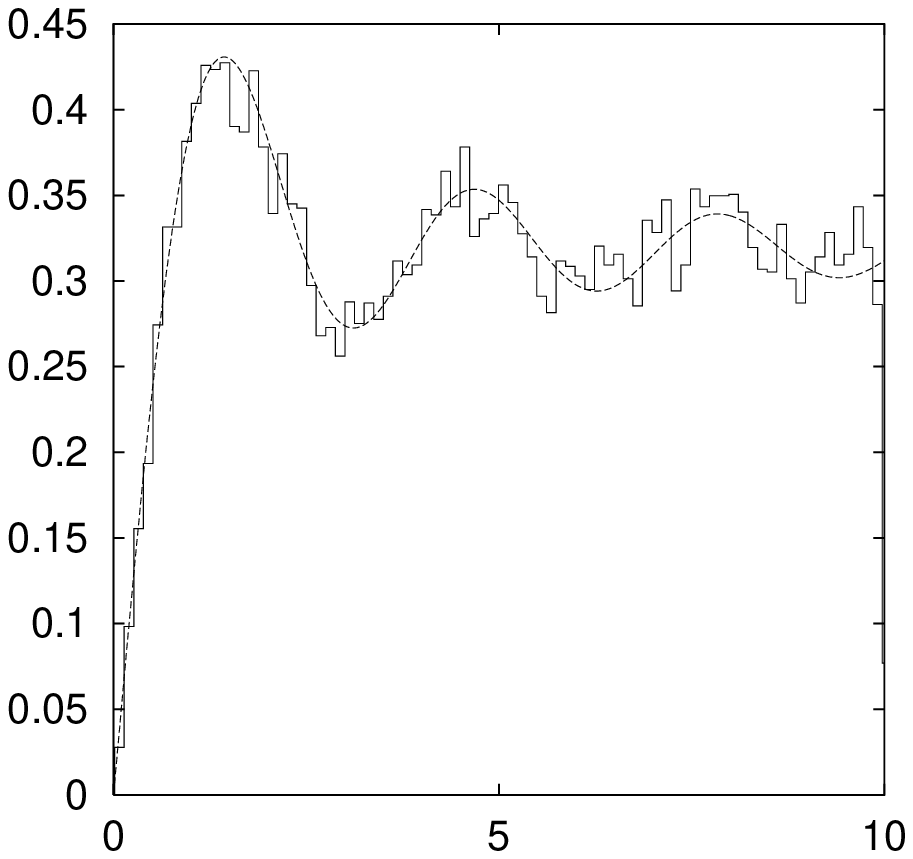,width=4.5cm}\\[-40mm]
    \hspace*{-50mm}$P(\lambda_{\min})$\hspace*{55mm}$\rho_s(z)$\\[35mm]
    \hspace*{38mm}$\lambda_{\min}$\hspace*{60mm}$z$
  \end{center}
  \caption{Distribution $P(\lambda_{\min})$ (left plot) and microscopic
    spectral density $\rho_s (z)$ (right plot) from our $6^4$ lattice data
    of U(1) gauge theory in comparison with the predictions of the chUE
    of RMT (dashed lines).}
  \label{f06}
\end{figure}

To study the smallest eigenvalues, spectral averaging is not possible,
and one has to produce large numbers of configurations. Our present
results are for $\beta=0.9$ in the confined phase with 10000
configurations on a $4^4$, 10000 configuration on a $6^4$, and 2822
configurations on an $8^3 \times 6$ lattice.  The left plot in
Fig.~\ref{f06} exhibits the distribution $P(\lambda_{\min})$ of the
smallest eigenvalue $\lambda_{\min}$ in comparison with the prediction
of the (quenched) chUE of RMT for topological charge $\nu=0$,
\begin{equation}
\label{plambdamin}
P(\lambda_{\min}) = {(V\Sigma)^2 \lambda_{\min} \over 2}\,\exp\left( - 
{(V\Sigma\lambda_{\min})^2 \over 4} \right).
\end{equation}
The agreement is excellent for all lattices.  For the chiral
condensate we obtain $\Sigma \approx 0.35$ by extrapolating the
histogram for $\rho(\lambda)$ to $\lambda=0$ and using the
Banks-Casher relation. 
Since the average value of $\lambda_{\min}$ goes like $V^{-1}$,
$\langle\lambda_{\min}\rangle$ decreases with increasing lattice size.
In the right plot of
Fig.~\ref{f06} the same comparison with RMT is done for the microscopic
spectral density $\rho_s (z)$ up to $z=10$, and the agreement is
again quite satisfactory. Here, the analytical RMT result for the
(quenched) chUE and $\nu=0$ is given by \cite{ShVe92}
$ \rho_s(z) = z\, [ J_0^2(z) + J_1^2(z) ]/2$, where $J$ denotes the
Bessel function.

\begin{figure}
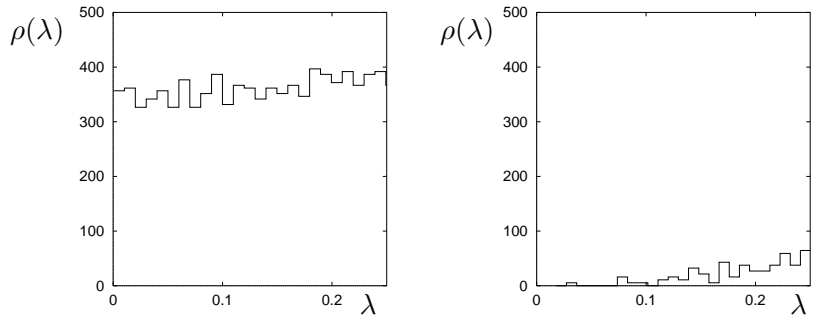

  \begin{center}
    \psfig{figure=f12.eps,width=4cm}\hspace*{15mm}
    \psfig{figure=f11.eps,width=4cm}\\[-40mm]
    \hspace*{-50mm}$\rho(\lambda)$\hspace*{50mm}$\rho(\lambda)$\\[32mm]
    \hspace*{38mm}$\lambda$\hspace*{55mm}$\lambda$
  \end{center}
  \caption{Density $\rho(\lambda)$ of small eigenvalues for the
    $8^3\times 6$ lattice at $\beta = 0.9$ (left plot) and at $\beta =
    1.1$ (right plot). A nonzero chiral condensate is supported in
    the confinement phase of U(1) gauge theory.}
  \label{f12}
\end{figure}

The quasi-zero modes which are responsible for the chiral condensate
$\Sigma \approx 0.35$ build up when we cross from the Coulomb into the
confined phase. For our $8^3\times 6$ lattice, Fig.~\ref{f12} compares
on identical scales densities of the small eigenvalues at $\beta =
0.9 $ (left plot) and at $\beta = 1.1$ (right plot), averaged over
20 configurations. 
The quasi-zero modes in the left plot are related to the nonzero
chiral condensate $\Sigma>0$, whereas no
such quasi-zero modes are found in the Coulomb phase. 
It may be worthwhile to understand the physical
origin of the U(1) quasi-zero modes in more detail. For 4d SU(2) and
SU(3) gauge theories a general interpretation is to link them, and
hence the chiral condensate, to the existence of instantons. As there
are no instantons in 4d U(1) gauge theory, one needs another
explanation,
and it is interesting to study local correlations of the
fermion density with the topological charge density and the monopole
density \cite{Sak99}.

\begin{figure}
  \begin{center}
    \psfig{figure=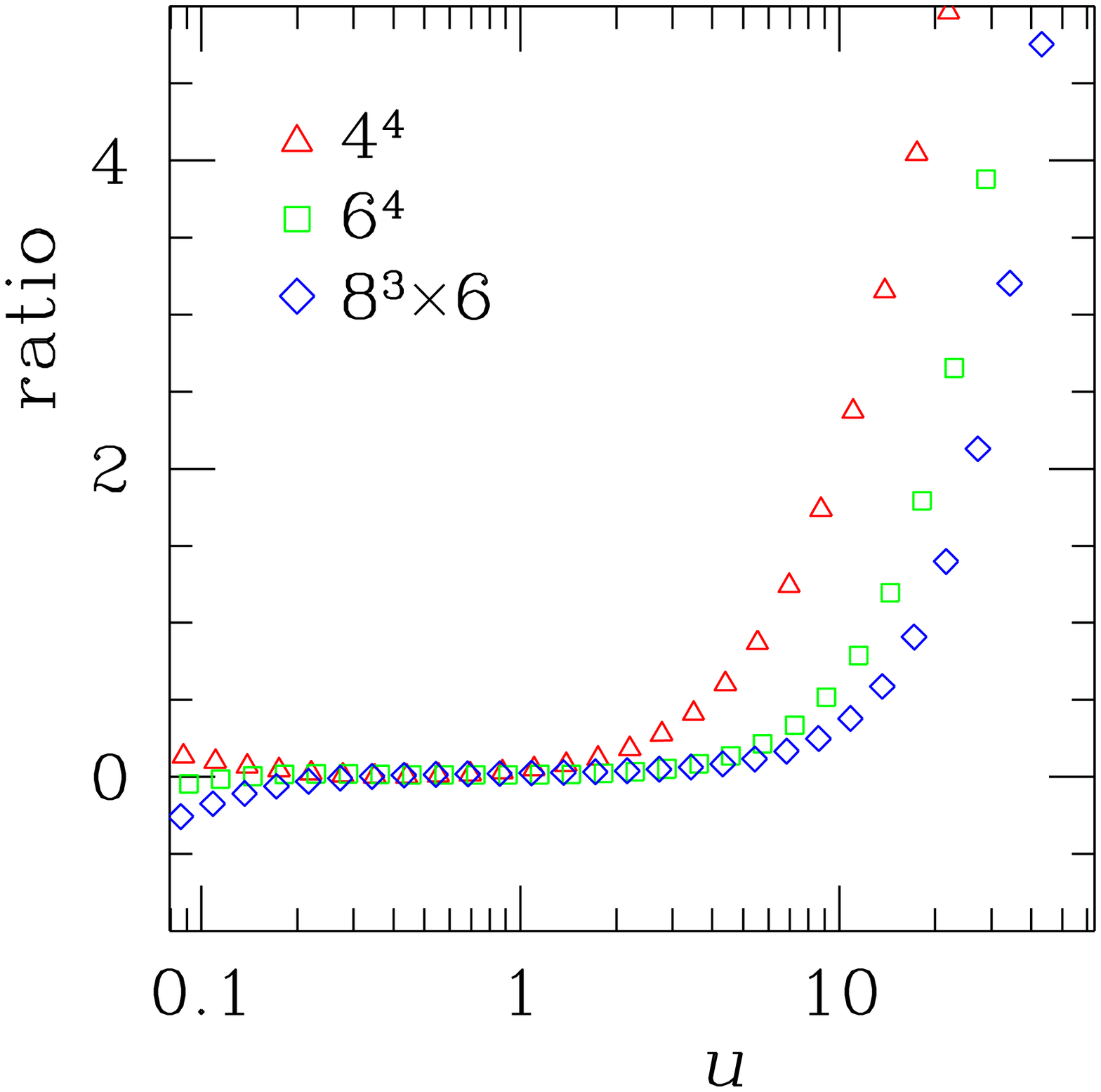,width=45mm}\hspace*{10mm}
    \psfig{figure=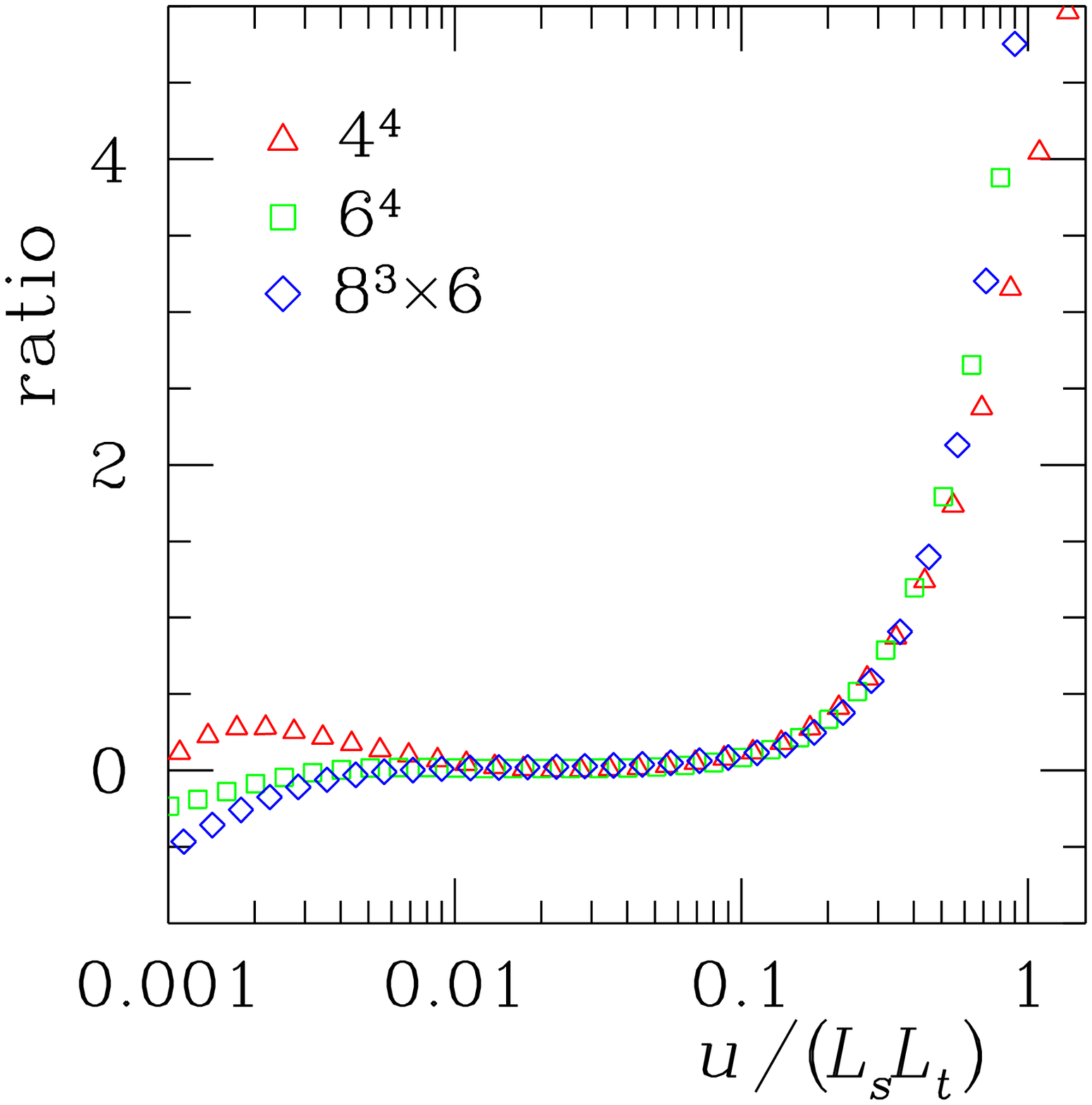,width=45mm}
    \caption{The ratio of Eq.~(\ref{ratio}) for U(1) gauge theory 
      plotted versus $u$ and $u/(L_sL_t)$, respectively (error bars
      not shown).  In the right plot, the data for different $L_s$ 
      and $L_t$ fall on the
      same curve, confirming the expected scaling of the Thouless
      energy according to Eq.~(\ref{thouless}).  The deviations of the
      ratio from zero for very small values of $u$ are well-understood
      artifacts of the finite lattice size and finite statistics
      \protect\cite{Berb98}.}
    \label{fig8}
  \end{center}
\end{figure}

Another interesting question concerns the energy scale $E_c$ up to
which RMT describes the small Dirac eigenvalues in the phase where
$\rho(0)>0$.  In disordered mesoscopic systems, a similar scale is
called the Thouless energy.  The theoretical prediction for QCD is
$E_c\sim f_\pi^2/\Sigma L^{2}_{s}$ \cite{Osbo98} with the pion decay
constant $f_{\pi}$, where we have assumed that the spatial extent
$L_s$ of the lattice is not smaller than the temporal extent $L_t$.
In units of the mean level spacing $\Delta=\pi/V\Sigma$ at the origin,
this becomes
\begin{equation}
  \label{thouless}
  u_c\equiv \frac{E_c}{\Delta} \sim \frac1\pi f_\pi^2L_sL_t\:.
\end{equation}
A convenient quantity from
which $u_c$ can be extracted is the disconnected scalar
susceptibility,
\begin{equation}
  \chi_{\rm latt}^{\rm disc}(m)=\frac1N\left\langle\sum_{k,l=1}^N
    \frac1{(i\lambda_k+m)(i\lambda_l+m)}\right\rangle_{\!\!A}
  -\frac1N\left\langle
    \sum_{k=1}^N\frac1{i\lambda_k+m}\right\rangle_{\!\!A}^2\:.
\end{equation}
The corresponding RMT result for the quenched chUE with $\nu=0$ reads
\cite{Goec99} $\chi_{\rm RMT}^{\rm
  disc}=u^2[K_1^2(u)-K_0^2(u)][I_0^2(u)-I_1^2(u)]$, where
$u=mV\Sigma$, and $I$ and $K$ are modified Bessel functions.  In
Fig.~\ref{fig8} we have plotted the ratio \cite{Berb98}
\begin{equation}
  \label{ratio}
  {\rm ratio}=\left(\chi_{\rm latt}^{\rm disc}-\chi_{\rm RMT}^{\rm
      disc}\right)/\chi_{\rm RMT}^{\rm disc}
\end{equation}
versus $u$ and $u/(L_sL_t)$, respectively, for the U(1) data computed
at $\beta=0.9$.  This ratio should deviate from zero above the
Thouless scale.  The expected scaling of the Thouless energy with
$L_sL_t$ is confirmed.

\section{Conclusions}

We have searched for a transition in the nearest-neighbor spacing
distribution $P(s)$ from Wigner to Poisson behavior across the
deconfinement phase transition of pure gluonic and of full QCD. We
observed no signature of a transition,
neither for pure SU(3) nor for full QCD.  The data agree
with the RMT result in both 
phases, except for extremely large values of $\beta$ where the
eigenvalues are known analytically. Our analysis of full QCD shows
that quark masses have no influence on the nearest-neighbor spacing
distribution. One explanation of our results is that temporal monopole
currents survive the deconfinement phase transition leading to
confinement of spatial Wilson loops.  Thus, even in the deconfinement
phase, the gauge fields retain a considerable degree of randomness.

A general unfolding procedure for the spectra of non-hermitian operators
was applied to the QCD lattice Dirac operator at nonzero chemical potential.
Agreement of the nearest-neighbor spacing distribution with predictions of
the Ginibre ensemble of non-hermitian RMT was found between $\mu=0.5$ and
$\mu=0.7$ in both confinement and deconfinement phase.  The deviations
from Ginibre behavior for smaller values of $\mu$ as well as the changes
for larger values of $\mu$ toward a Poisson shape are understood mathematically.
The physical interpretation requires a better understanding of QCD at nonzero
density. 
An interesting observation is that the results for
$P(s)$ in the non-hermitian case are rather sensitive to $\mu$ whereas
they are very stable under variations of $T$ in the hermitian case.

The nearest-neighbor spacing distribution of 4d U(1) quenched lattice
gauge theory is described by the chUE in both the confinement and the
Coulomb phase.  In the confinement phase we also find that the
$P(\lambda_{\min})$ distribution and the microscopic spectral density
$\rho_s(z)$ are described by the chUE.  The Thouless energy
scales with the lattice size as expected.


\section{Acknowledgments}

This work was supported in part by FWF projects P10468-PHY and
P11456-PHY, by DFG grants We 655/11-2 and We 655/15-1, by DOE
contracts DE-FG02-97ER41022, DE-FG05-85ER2500, and DE-FG02-91ER40608,
and by the RIKEN BNL Research Center.  We thank T.S.\ Bir\'o, E.-M.\ 
Ilgenfritz, N. Kaiser, M.I.\ Polikarpov, K. Rabitsch, and J.J.M.\ 
Ver\-baar\-schot for helpful discussions.


\end{document}